\documentclass[a4paper]{article}

\pdfoutput=1

\usepackage{amsmath}
\usepackage{amssymb}
\usepackage{amstext}
\usepackage{graphicx}
\usepackage{fullpage}
\usepackage{xcolor}
\usepackage{setspace}

\usepackage[skins,theorems]{tcolorbox}
\usepackage{authblk}

\providecommand{\keywords}[1]
{
  \small	
  \textbf{\textit{Keywords---}} #1
}

\begin{document}

\title{Finite element solutions of the nonlinear RAPM Black-Scholes model}

\author[1,*]{Dongming Wei}
\author[2,*]{Yogi Ahmad Erlangga}
\author[1,3]{Andrey Pak}
\author[1]{Laila Zhexembay}
\affil[1]{Nazarbayev University, Department of Mathematics, School of Sciences and Humanities, 53 Kabanbay Batyr Ave, Nur-Sultan 010000, Kazakhstan}
\affil[2]{Zayed University, Department of Mathematics, College of Natural and Health Sciences, Abu Dhabi Campus, P.O. Box 144534, United Arab Emirates}
\affil[3]{University of Alberta, Faculty of Science, Department of  Mathematics and  Statistical Sciences, Edmonton AB T6G 2G1,
Canada}
\affil[ ]{\textit{dongming.wei@nu.edu.kz, yogi.erlangga@zu.ac.ae, pak@ualberta.ca}}
\affil[*]{Corresponding author}

\date{ }
\maketitle

\begin{abstract} This paper presents finite element methods for solving numerically  the Risk-Adjusted Pricing Methodology (RAPM) Black-Scholes model for option pricing with transaction costs. Spatial finite element models based on P1 and/or P2 elements are formulated using some group finite elements and numerical quadrature to handle the nonlinear term, in combination with a Crank-Nicolson-type temporal scheme. The temporal scheme is implemented using the Rannacher approach. Spatial-temporal mesh-size ratios are observed for controlling the stability of our method.  Our results compare favorably with the finite difference results in the literature for the model.
\end{abstract}


\section{Introduction} \label{sec:intro}

A fair option price in a complete financial market with no transaction costs can be modelled by the Black-Scholes equation~\cite{scholesB73, merton73}. The underlying assumption requires, however, that portfolio hedging takes place continuously. In the market with transaction costs, this assumption becomes unrealistically expensive. Modifications to the Black-Scholes (BS)  model have been proposed to count for the transaction costs, which lead to various nonlinear models~\cite{leland85, Barles98S, Kratka98, Hoggard94WW, Janda05S}. Recent overview of nonlinear extensions to the Black–Scholes option pricing models is presented in \cite{vsevcovic2011analytical} (see Ch. 11). 

The object of the study in the current paper is the Risk-Adjusted Pricing Methodology (RAPM) model, that was first introduced in \cite{Kratka98}, and subsequently improved in \cite{Janda05S}. This model incorporates both transaction costs and the risks arising from a volatile portfolio. By minimizing the total risk premium, which is the sum of transaction costs and the risk cost from the unprotected volatile portfolio, Janda{\v{c}}ka and {\v{S}}ev{\v{c}}ovi{\v{c}}\cite{Janda05S} found the optimal length of the hedge interval. Consequently, the authors were able to obtain new strategies for hedging options, which are associated with a solution to the nonlinear parabolic BS equation with a diffusion coefficient nonlinearly depending on the option price itself. More precisely, for an European call option with the strike price $K$ and expiration time $T$, its price $V(S,t)$ at time $t\in [0,T]$ can then be modelled in the following way:

(a) On the time interval $(0,t_*), V(S,t)$ is governed by the nonlinear partial differential equation:
\begin{eqnarray}
V_t + \frac{1}{2}\tilde{\sigma}^2 S^2 V_{SS} + r SV_S - r V = 0, \quad \text{in } (0,t_*) \times \mathbb{R}_+ \label{eq:BSRAPM}
\end{eqnarray}
where
\begin{align}
\displaystyle \tilde{\sigma}^2 = \sigma^2 \left( 1 + 3 \left[\frac{C^2 M}{2\pi} S V_{SS} \right]^{\frac{1}{3}} \right), \label{eq:BSRAM1}
\end{align}
$S$ is the value of the underlying asset,  $r$ is the risk-free interest rate, $\sigma$ is the volatility, $M \ge 0$ is the transaction cost measure, and $C\ge 0$ is the risk premium measure. The switching time $t_*$ is a time of the very last portfolio adjustment before the expiry time $T,$ and is calculated by $t_*=T-C/(M\sigma^2)$. For~\eqref{eq:BSRAPM}, in addition to the boundary conditions
\begin{eqnarray}
  V(0,t) &=& 0, \label{bcond0} \\
  V(S,t) &\sim& S - Ke^{-r(T-t)}, \text { as } S \to \infty, \label{bcond1} 
\end{eqnarray}
for all $t \in [0,t_*]$, the following terminal condition at the switching time $t_*$ is  required:
\begin{eqnarray}
   V(S,t_*) = S\Phi(d_1)-Ke^{-r(T-t_*)}\Phi(d_2), \quad S \ge 0, \label{Tcond}
\end{eqnarray}
where $\Phi(x)$ is the cumulative distribution function of a standard normal distribution and
\begin{align*}
    d_1=\frac{\ln (S/K)+(r+\sigma^2/2)(T-t_*)}{\sigma \sqrt{T-t_*}}, \quad d_2=d_1-\sigma\sqrt{T-t_*}.
\end{align*}

(b) On the time interval $(t_*, T),\ V(S,t)$ obeys the classical Black-Scholes equation
\begin{eqnarray}
V_t + \frac{1}{2}\sigma^2 S^2 V_{SS} + r SV_S - r V = 0, \quad \text{in } (t^*,T) \times \mathbb{R}_+ \label{eq:BS}
\end{eqnarray}
satisfying the terminal condition at the expiration time $T:$
\begin{eqnarray}
   V(S,T) = \max(S-K,0), \quad S \ge 0. \label{icondT}
\end{eqnarray}
The condition~\eqref{icondT} is referred to as the pay-off function in literature. 

The solution to the \eqref{eq:BS}-\eqref{icondT} is the classical Black-Scholes formula, which is obtained by replacing $t_*$ with $t\in (t_*,T)$ in \eqref{Tcond}. As a result, in this paper we focus our attention on the solution of \eqref{eq:BSRAPM}. By assuming conditions $C<\sigma^2MT$ and $CM < \pi/8,$ we guarantee that $t_*$ is well-defined and solution of \eqref{eq:BSRAPM} exists (see \cite{Janda05S}). 

Practical option pricing is typically done by solving the underlying terminal-boundary value problem numerically. Popular numerical methods for this purpose are based on finite difference methods (FDM) and finite element (FEM) methods~\cite{Achdou05P}. FDM are particularly popular for both linear and nonlinear cases due to the simplicity of the methods, especially when the computation is performed on a uniform mesh~\cite{AnkudinovaE08,Company09JP, Zhao16YW}. Development of high-order methods as well as mesh adaptivity used to control numerical errors may however not be trivially done with FDM~\cite{Duering03FJ,liaoK09, Linde09PS, Gulen19PS}. These are not an issue with FEM, even though the implementation is more complex than FDM~\cite{Piron99H}. While FEM have been demonstrated to be a viable alternative to FDM in the linear cases~\cite{Marko08, Anda11AS, Golba13BA}, only limited work is presently done on the nonlinear cases, especially involving transaction costs under Leland's model (see~\cite{Almeida17CD}).

In~\cite{Wei20EZ} we demonstrate implementation of finite element methods to solve the nonlinear Black-Scholes equation based on Leland's transaction cost model. As a natural extension of ~\cite{Wei20EZ}, in this paper, we present some novel finite element models and algorithms with simulation results of the European option pricing based on solving numerically the nonlinear RAPM model~\eqref{eq:BSRAPM}.

The remainder of the paper is organized as follows. After introducing transformation of the RAPM model~\eqref{eq:BSRAPM} into a more convenient form in Section~\ref{sec:RAPMtrans}, we discuss a finite-element method and treatment for the nonlinear term in Section~\ref{sec:FEM}.  Our approach in tackling the nonlinearity based on  group finite elements and numerical quadrature results in  a novel finite element model.  Section~\ref{sec:time} discusses the time-integration method using  the  Rannacher  approach. Numerical results from the simulations are presented in Section~\ref{sec:result}, followed by concluding remarks in Section~\ref{sec:conclusion}.

\section{The RAPM model} \label{sec:RAPMtrans}

To solve the nonlinear RAPM Black-Scholes equation, we first transform the equation using the following changes of variables:
\begin{align}
  x = \ln(S/K), \quad \tau = \frac{1}{2}\sigma^2 (T-t), \quad u(x,\tau) = e^{-x} V/K.  \label{eq:changevar}
\end{align}
The derivatives of $V$ can then be related to the derivatives of $u$ as follows:
\begin{align}
  V_t = -\frac{1}{2}\sigma^2 Su_{\tau}, \quad V_S = u_x + u, \quad \text{and} \quad V_{SS} = \frac{1}{S}(u_{xx}+u_x).
\end{align}
Substitution of the above derivatives to~\eqref{eq:BSRAPM} results in
\begin{align}
    u_{\tau} - \left( 1 + 3 \left[ \frac{C^2 M}{2\pi} (u_{xx} + u_x) \right]^{\frac{1}{3}} \right)  (u_{xx} + u_x) - Du_x  = 0, \quad \text{in } \mathbb{R} \times \mathbb{R}_+, \label{eq:BSRAPMtrans}
\end{align}
where $D = 2r/\sigma^2$. The change of variables in ~\eqref{eq:changevar} also transform the terminal and boundary conditions \eqref{bcond0}--\eqref{Tcond} to
\begin{align}
   u(x,\tau_*) &= \Phi(d_1)-e^{-(D\tau_*+x)}\Phi(d_2), \quad x \in \mathbb{R}, \\
   u(x,\tau) &=0, \quad \text{as } x \to -\infty, \\
   u(x,\tau) &= 1 - e^{-D\tau - x}, \quad \text{as } x \to \infty,
\end{align}
for $\tau \in [\tau_*, \frac{1}{2}\sigma^2 T],$ where $\tau_*=\frac{C}{2M}$, $ d_1=\frac{x+ (D+1)\tau_* }{\sqrt{2\tau_*}}$, and $d_2 = d_1-\sqrt{2\tau_*}.$

For computational purposes, we truncate the solution domain to $\Omega = [-R,R]$, where $0 < R < \infty$ and $R$ is taken to be a large number to approximate the boundary condition at $x\to \infty$. We enforce the condition at $-\infty$ to be satisfied at $x = -R$, and similarly for the other boundary condition.

\section{Finite elements for the RAPM model} \label{sec:FEM}

To construct our finite element model for approximations to~\eqref{eq:BSRAPMtrans}, we first rewrite the PDE as a mixed formulation
\begin{align}
    &u_{\tau} - v - D u_x - C_{R}v^{4/3}  = 0, \label{eq:RAPMu} \\
   &v = u_{xx} + u_x
\end{align}
where $C_R = 3 \sqrt[3]{C^2M/2\pi}$. For Galerkin's finite element method, we consider the weak formulation with the test function $w$ and $z$:
\begin{eqnarray}
   \int\displaylimits_{\Omega} w \left(u_{\tau} - v - D u_x - C_{R}v^{4/3}\right) dx &=& 0, \notag \\
   \int\displaylimits_{\Omega} z (v - (u_{xx} + u_x)) d x &=& 0, \notag
\end{eqnarray}
which, after integration by parts, can be written as
\begin{eqnarray}
  \frac{\partial}{\partial\tau} \int\displaylimits_{\Omega} w u dx - \int\displaylimits_{\Omega} \left(wv +Dwu_x + C_R wv^{\frac{4}{3}}\right) dx &=& 0, \label{weak1} \\
   \int\displaylimits_{\Omega} z v dx  + \int\displaylimits_{\Omega} z_x u_x dx - \int\displaylimits_{\Omega} z u_x dx &=& 0.  \label{weak2}
\end{eqnarray}

Let $\displaystyle u = \sum_{i=1}^n u_i \psi_i + \sum_{i \in \mathcal{I}_{\partial \Omega}} u_i \psi_i$, $\mathcal{I}_{\partial \Omega} =\{0,n+1\}$, be the finite element approximation of the solution $u$,  where the second sum is the extension of the solution to the boundary $\partial \Omega=\{-R,R\}$ and $\psi_i$ is the global finite element shape function for the $i^{th}$ node in a spatial division $-R=x_0<...<x_i<...<x_{n+1}=R$. Similarly, we have  $\displaystyle v = \sum_{i=0}^{n+1} v_i \phi_i$, in which  no boundary conditions are set for $v$. Then \eqref{weak1} can be written as
\begin{align}
  0 &= \frac{\partial}{\partial \tau} \sum_{i=1}^n u_i \int\displaylimits_{\Omega} w \psi_i dx  - \sum_{i=1}^n v_i \int_{\Omega} w \phi_i dx - D \sum_{i=1}^n u_i \int_{\Omega} w \psi_{i,x} dx - C_R \int_{\Omega} wv^{\frac{4}{3}} dx \notag \\
    &+ \frac{\partial}{\partial \tau}  \sum_{i \in \mathcal{I}_{\partial \Omega}} u_i \int\displaylimits_{\Omega} w \psi_i dx - D \sum_{i \in \mathcal{I}_{\partial \Omega}} u_i \int_{\Omega} w \psi_{i,x} dx.
\end{align}
Enforcing this condition to be satisfied by $n$ functions $w_j$, $j=1,\dots,n$ yields a system of $n$ equations
\begin{align}
0 &= \frac{\partial}{\partial \tau} \sum_{i=1}^n u_i \int\displaylimits_{\Omega} w_j \psi_i dx  - \sum_{i=1}^n v_i \int_{\Omega} w_j \phi_i dx - D \sum_{i=1}^n u_i \int_{\Omega} w_j \psi_{i,x} dx - C_R \int_{\Omega} w_j v^{\frac{4}{3}} dx \notag \\
    &+ \frac{\partial}{\partial \tau}  \sum_{i \in \mathcal{I}_{\partial \Omega}} u_i \int\displaylimits_{\Omega} w_j \psi_i dx - D \sum_{i \in \mathcal{I}_{\partial \Omega}} u_i \int_{\Omega} w_j \psi_{i,x} dx. \label{eq:weak1a}
\end{align}
Similarly, for \eqref{weak2}, after enforcing the above equation to be satisfied by $z_j$, $j = 1, \dots, n$ results in the system of $n$ equations
\begin{align}
0 &= \sum_{i=1}^{n} v_i  \int\displaylimits_{\Omega} z_j \phi_i dx +  \sum_{i=1}^n u_i \left\{ \int \displaylimits_{\Omega} z_{j,x} \psi_{i,x} dx - \int \displaylimits_{\Omega} z_j \psi_{i,x} dx \right\} \notag \\
&+  \sum_{i \in \mathcal{I}_{\partial \Omega}} u_i \left\{  \int \displaylimits_{\Omega} z_{j,x} \psi_{i,x} dx - \int \displaylimits_{\Omega} z_j \psi_{i,x} dx    \right\}.  \label{eq:weak2a}
\end{align}
Considering the Galerkin approach with $w_i = z_i = \phi_i = \psi_i$, Equations~\eqref{eq:weak1a} and~\eqref{eq:weak2a} then become, for $j=1,\dots, n$,
\begin{align}
 \frac{\partial}{\partial \tau} \sum_{i=1}^n u_i \int\displaylimits_{\Omega} \psi_j \psi_i dx  &- \sum_{i=1}^n v_i \int_{\Omega} \psi_j \psi_i dx - D \sum_{i=1}^n u_i \int_{\Omega} \psi_j \psi_{i,x} dx - C_R \int_{\Omega} \psi_j v^{\frac{4}{3}} dx \notag \\
 &= -\frac{\partial }{\partial \tau}  \sum_{i \in \mathcal{I}_{\partial \Omega}} u_i \int\displaylimits_{\Omega} \psi_j \psi_i dx 
    +  D \sum_{i \in \mathcal{I}_{\partial \Omega}} u_i \int_{\Omega} \psi_j \psi_{i,x} dx, \label{eq:galerkin1a} \\
\sum_{i=1}^{n} v_i  \int\displaylimits_{\Omega} \psi_j \psi_i dx &+  \sum_{i=1}^n u_i \left\{ \int \displaylimits_{\Omega} \psi_{j,x} \psi_{i,x} dx - \int \displaylimits_{\Omega} \psi_j \psi_{i,x} dx \right\} \notag \\
&=  - \sum_{i \in \mathcal{I}_{\partial \Omega}} u_i \left\{  \int \displaylimits_{\Omega} \psi_{j,x} \psi_{i,x} dx - \int \displaylimits_{\Omega} \psi_j \psi_{i,x} dx    \right\}.  \label{eq:galerkin2a}
\end{align}
Let the domain $\Omega$ be subdivided into $n_E$ nonoverlapping elements such that $\Omega = \bigcup\limits_{i=1}^{n+1} \Omega_{i}$, where $\Omega_{i}= [x_{i-1},x_i]$, the $i$-th element with boundary nodes $x_{i-1}$ and $x_i$. In this way, each integral above can be written as the sum of integral over each element. For instance
$$
   \int \displaylimits_\Omega \psi_j \psi_i dx = \sum_{\ell=1}^n \int \displaylimits_{\Omega_\ell} \psi_j \psi_i dx,
$$
and so on. Thus, the integral over the domain $\Omega$ can be evaluated by first evaluating integral over elements and then summing up, a process referred to as ``assembly''. In the implementation, the assembly process is based on element matrices that represents integral terms in~\eqref{eq:galerkin1a} and~\eqref{eq:galerkin2a} over each element $\Omega_j$. Structures of the element matrices depend on the specific choice of the functions $\psi_i$.
Specifically, the interpolation functions $\psi_i$ are chosen such that, at the nodal points $x_j$,
\begin{eqnarray}
    \psi_i(x_j) = \begin{cases}
                         1,& i = j, \\
                         0,&\text{otherwise}.
                    \end{cases}
\end{eqnarray}
In this way, at the left boundary point $x_0 = -R$,
$$
  u(x_0) = \sum_{i=n} u_i \psi(x_0) + \sum_{i \in \mathcal{I}_{\partial \Omega}} u_i \psi(x_0) = u_0 = 0.
$$
Similarly at the right boundary point $x_n = R$, 
$$
  u(x_n) = \sum_{i=n} u_i \psi(x_n) + \sum_{i \in \mathcal{I}_{\partial \Omega}} u_i \psi(x_n) = u_n = 1 - e^{-D\tau - R}.
$$

In the sequel, we shall discuss only treatment for the nonlinear part; for the treatment of linear parts, see, e.g., ~\cite{Wei20EZ, Aichinger13B}.

\subsection{Group finite elements}

Let $f(v) = v^{\frac{4}{3}}$ and consider the approximation:
\begin{align}
   f(v) = \sum_{i=1}^n f_i(v) \psi_i, \quad f_i(v) = f(v(x_i)).
\end{align}
Therefore,
\begin{align}
   \int_{\Omega} \psi_j v^{\frac{4}{3}} dx \simeq \int_{\Omega} \psi_j \sum_{i=1}^n f_i(v) \psi_i dx = \sum_{i=1}^n f_i(v) \int_{\Omega} \psi_j \psi_i dx = \sum_{i=1}^n v_i^{\frac{4}{3}} \int_{\Omega} \psi_j \psi_i dx.
\end{align}

\subsection{Numerical quadrature}

In this approach, by setting $f(x) = \psi_j v^{\frac{4}{3}}$,  the integral $\displaystyle \int_{\Omega} f(x)dx$ is evaluated approximately using some numerical quadrature.

In the $\Omega_j$ element, using the trapezoidal rule,
\begin{align}
   \int_{\Omega_j} \psi_{j-1} v^{\frac{4}{3}}dx &= \int_{\Omega_j} \psi_{j-1} \left( v_{j-1} \psi_{j-1} + v_j \psi_j \right)^{\frac{4}{3}} dx \notag \\
   &= \frac{h_j}{2} [ \psi_{j-1}(x_{j-1}) (\left( v_{j-1} \psi_{j-1}(x_{j-1}) + v_j \psi_j (x_{j-1}) \right)^{\frac{4}{3}} + \psi_{j-1} (x_j) \left( v_{j-1} \psi_{j-1}(x_j) + v_j \psi_j(x_j) \right)^{\frac{4}{3}} ] \notag \\
   &= \frac{1}{2}h_jv_{j-1}^{\frac{4}{3}}, \notag
\end{align}
because $\psi_{j-1}(x_{j-1}) = \psi_j(x_j) = 1$ and $\psi_{j-1}(x_j) = \psi_j(x_{j-1}) = 0$. Similarly,
\begin{align}
   \int_{\Omega_j} \psi_j v^{\frac{4}{3}}dx &= \int_{\Omega_j} \psi_j \left( v_{j-1} \psi_{j-1} + v_j \psi_j \right)^{\frac{4}{3}} dx \notag \\
   &= \frac{h_j}{2} [ \psi_j(x_{j-1}) (\left( v_{j-1} \psi_{j-1}(x_{j-1}) + v_j \psi_j (x_{j-1}) \right)^{\frac{4}{3}} + \psi_j (x_j) \left( v_{j-1} \psi_{j-1}(x_j) + v_j \psi_j(x_j) \right)^{\frac{4}{3}} ] \notag \\
   &= \frac{1}{2}h_jv_j^{\frac{4}{3}}. \notag
\end{align}
The element matrix for the nonlinear term is given by
\begin{align}
  N_1^{(j)}[v] = \frac{h_j}{2}
  \begin{bmatrix}
     1 & 0 \\
     0 & 1
  \end{bmatrix} \begin{bmatrix}
     v_{j-1}^{\frac{4}{3}} \\
     v_j^{\frac{4}{3}}
  \end{bmatrix}
\end{align}

For the P2-element, we consider Simpsons's rule to approximate the integral.
\begin{align}
   \int_{\Omega_j} \psi_{j-1}v^{\frac{4}{3}} dx &= \int_{\Omega_j} \psi_{j-1} (v_{j-1} \psi_{j-1} + v_{j-\frac{1}{2}} \psi_{j-\frac{1}{2}} + v_j \psi_j)^{\frac{4}{3}} ) dx \notag \\
    &= \frac{h_j}{6} [ \psi_{j-1}(x_{j-1}) (v_{j-1} \psi_{j-1}(x_{j-1}) + v_{j-\frac{1}{2}} \psi_{j-\frac{1}{2}}(x_{j-1}) + v_j \psi_j(x_{j-1}))^{\frac{4}{3}} \notag \\
     & \quad \quad \quad + 4 \psi_{j-1}(x_{j-\frac{1}{2}}) (v_{j-1} \psi_{j-1}(x_{j-\frac{1}{2}}) + v_{j-\frac{1}{2}} \psi_{j-\frac{1}{2}}(x_{j-\frac{1}{2}}) + v_j \psi_j(x_{j-\frac{1}{2}}))^{\frac{4}{3}} \notag \\
     &  \quad \quad \quad + \psi_{j-1}(x_j) (v_{j-1} \psi_{j-1}(x_j) + v_{j-\frac{1}{2}} \psi_{j-\frac{1}{2}}(x_j) + v_j \psi_j(x_j))^{\frac{4}{3}} ] \notag \\
     &= \frac{h_j}{6} v_{j-1}^{\frac{4}{3}}. \notag
\end{align}
In a similar vein, we can show that
\begin{align}
   \int_{\Omega_j} \psi_{j-\frac{1}{2}} v^{\frac{4}{3}} dx &= \frac{4h_j}{6} v_{j-\frac{1}{2}}^{\frac{4}{3}},  \notag \\
   \int_{\Omega_j} \psi_j v^{\frac{4}{3}} dx &= \frac{h_j}{6} v_j^{\frac{4}{3}},  \notag
\end{align}
The element matrix for the nonlinear term is given by
\begin{align}
  N^{(j)}_2[v] = \frac{h_j}{6}
  \begin{bmatrix}
     1 & 0 & 0 \\
     0 & 4 & 0 \\
     0 & 0 & 1
  \end{bmatrix} \begin{bmatrix}
     v_{j-1}^{\frac{4}{3}} \\
     v_{j-\frac{1}{2}}^{\frac{4}{3}} \\
     v_j^{\frac{4}{3}}
  \end{bmatrix}.
\end{align}

\section{Time integration} \label{sec:time}

The global finite element system can be assembled using the element matrices derived in Section~\ref{sec:FEM} (see, e.g., ~\cite{ Aichinger13B}), resulting in the first-order nonlinear differential algebraic system of two equations:
\begin{align}
  \begin{cases}
    \frac{\partial }{\partial \tau} (M\mathbf{u} + \mathbf{b}_M)  = M \mathbf{v} + \varepsilon P \mathbf{u} +  N \mathbf{v}^{\frac{4}{3}} + \varepsilon \mathbf{b}_P, \\
    M\mathbf{v} + K \mathbf{u} - P \mathbf{u} = - (\mathbf{b}_K - \mathbf{b}_P).
  \end{cases}
\end{align}
Substitution of the second equation to the first results in 
\begin{align}
   \frac{\partial }{\partial \tau} (M\mathbf{u} + \mathbf{b}_M)  = - K \mathbf{u} +  (1 + \varepsilon) P \mathbf{u} +  N \mathbf{v}^{\frac{4}{3}} - \mathbf{b}_K + (1+ \varepsilon) \mathbf{b}_P =: \mathbf{F}(\mathbf{u}),
\end{align}
with $\mathbf{v}  = M^{-1}(-K \mathbf{u} + P \mathbf{u} - \mathbf{b}_K +  \mathbf{b}_P)$.

Integration over time is approximated using the Crank-Nicolson-type scheme:
$$
   \frac{1}{\Delta \tau} (M\mathbf{u}^{n+1} + \mathbf{b}_M^{n+1} - (M\mathbf{u}^{n} + \mathbf{b}_M^n)) = \theta \mathbf{F}^{n+1} + (1-\theta) \mathbf{F}^{n},
$$
with $\theta \in [0,1]$. $\theta = 0$ and 1 correspond to the explicit forward and implicit backward Euler method, respectively. Rearranging the above equation leads to
$$
   M\mathbf{u}^{n+1} - \theta  \Delta \tau \mathbf{F}^{n+1} =  M \mathbf{u}^n + (1-\theta) \Delta \tau \mathbf{F}^{n} - \mathbf{b}_M^{n+1} +  \mathbf{b}_M^n.
$$
One possible linearization can be based on the approximation
$$
  \mathbf{F}^{n+1} \simeq  - K \mathbf{u}^{n+1} +  (1 + \varepsilon) P \mathbf{u}^{n+1} +  N \text{diag}((\mathbf{v}^n)^{\frac{1}{3}}) \mathbf{v}^{n+1}- \mathbf{b}_K^{n+1} + (1+ \varepsilon) \mathbf{b}_P^{n+1}, 
$$
with $\mathbf{v}^{n+1} = M^{-1}(-K \mathbf{u}^{n+1} + P \mathbf{u}^{n+1} - \mathbf{b}_K^{n+1} +  \mathbf{b}_P^{n+1})$. Thence, we obtain the following time integration algorithm
\begin{align}
   A \mathbf{u}^{n+1} = M \mathbf{u}^n + (1-\theta) \Delta \tau \mathbf{F}^{n} - \mathbf{b}_M^{n+1} +  \mathbf{b}_M^n - N\text{diag}((\mathbf{v}^n)^{\frac{1}{3}}) (-\mathbf{b}_K^{n+1} +  \mathbf{b}_P^{n+1}),
\end{align}
with
$$
    A = M - \theta \Delta \tau (-K + (1+\varepsilon)P + N \text{diag}((\mathbf{v}^n)^{\frac{1}{3}}) (M^{-1}(-K + P))),
$$
and
$$
   \mathbf{F}^n = - K \mathbf{u}^n +  (1 + \varepsilon) P \mathbf{u}^n +  N (\mathbf{v}^n)^{\frac{4}{3}} - \mathbf{b}_K^n + (1+ \varepsilon) \mathbf{b}_P^n.
$$

For improved the stability of (30), the standard Crank-Nicolson scheme is implemented using the Rannacher approach~\cite{Rann84, Giles06C}, in which the first Crank-Nicolson step is replaced by a few backward implicit Euler steps with smaller time steps (e.g., $\Delta \tau_R = \Delta \tau/n_R$), where $n_R$ is the number of backward Euler time steps in from $\tau = 0$ to $\tau = \Delta \tau$).

\section{Numerical results} \label{sec:result}

We performed numerical simulations using the FEM model and the time-integration method discussed in Sections~\ref{sec:FEM} and~\ref{sec:time}. All results are computed on a uniform finite-element mesh, even though the method can be implemented on a nonuniform mesh. In all computations, we set the parameters in the RAPM model~\eqref{eq:BSRAPM} and~\eqref{eq:BSRAM1} as follows: $r = 0.1$, $\sigma = 0.2$, $C = 0.01$, $M = 2$, $T = 1$, and $J = 75$. Under the setting, the condition for the existence of a solution is satisfied.

\begin{figure}[!h]
\centering
\includegraphics[width=0.45\textwidth]{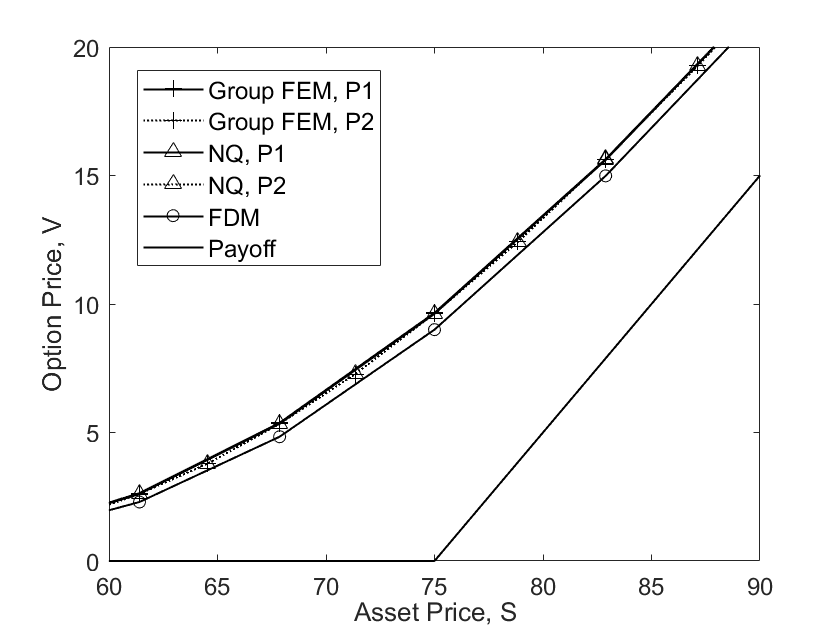}
\includegraphics[width=0.45\textwidth]{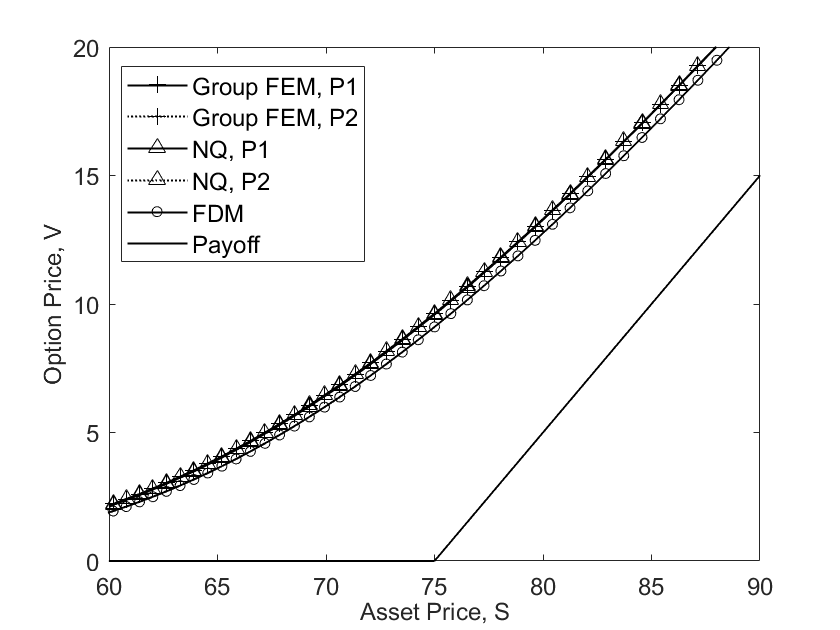}
\caption{Solution of the RAPM model, with $r = 0.1$, $\sigma = 0.2$, $K = 75$, $T = 1$, $C = 0.01$ and $M = 2$. Left figure: $\Delta \tau = 0.0005$, $\Delta x = 0.01$; Right figure: $\Delta \tau = 0.001$, $\Delta x = 0.001$.}
\label{fig:femvsfdm}
\end{figure}

Figure~\ref{fig:femvsfdm} shows solutions of FEM at $T = 0$ ($t = 1$ towards expiration), which lie close to each others. This result suggests that the simple-to-implement group finite element method works as effective as the more-complicated finite-element method with numerical quadrature. The FEM results also compare favorably with the second-order finite-difference method, described in~\cite{AnkudinovaE08}.

\begin{figure}[!h]
\centering
\includegraphics[width=0.45\textwidth]{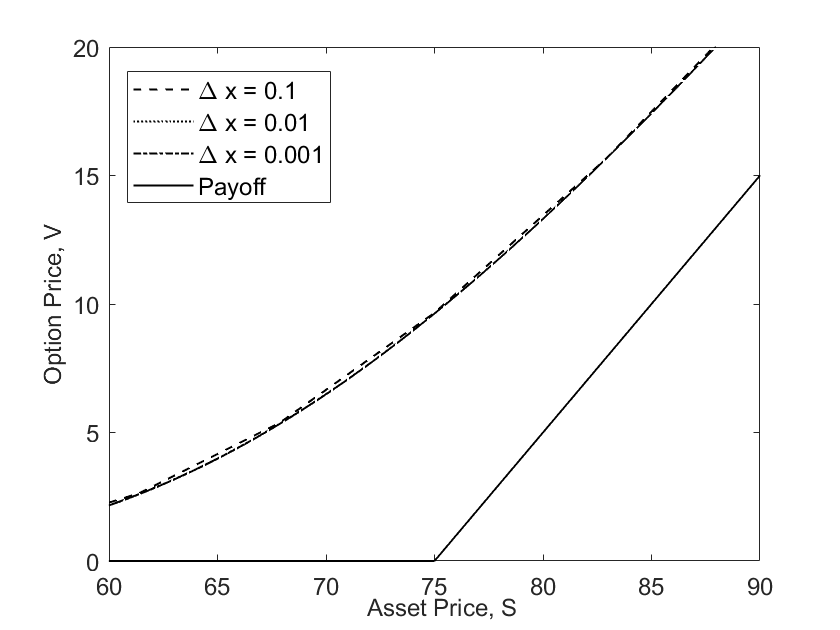}
\includegraphics[width=0.45\textwidth]{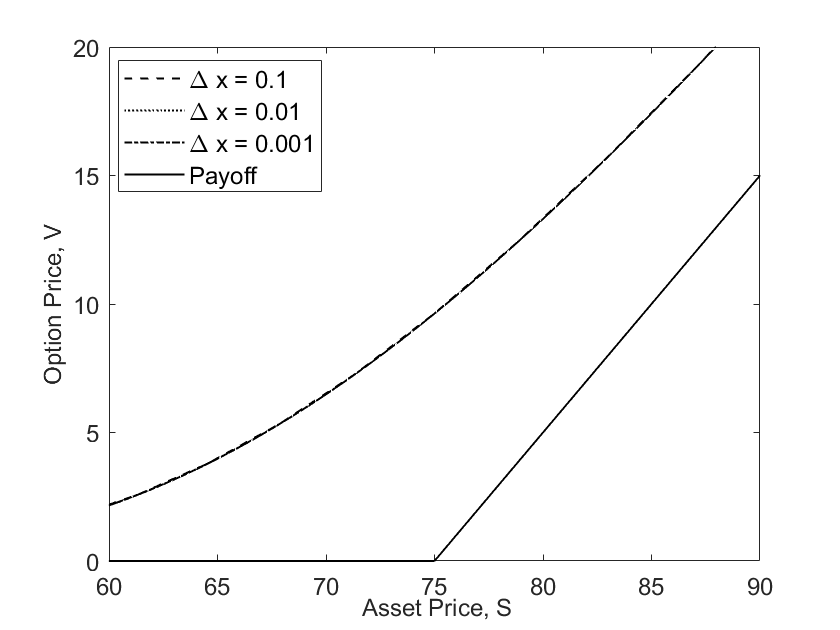} \\
\includegraphics[width=0.45\textwidth]{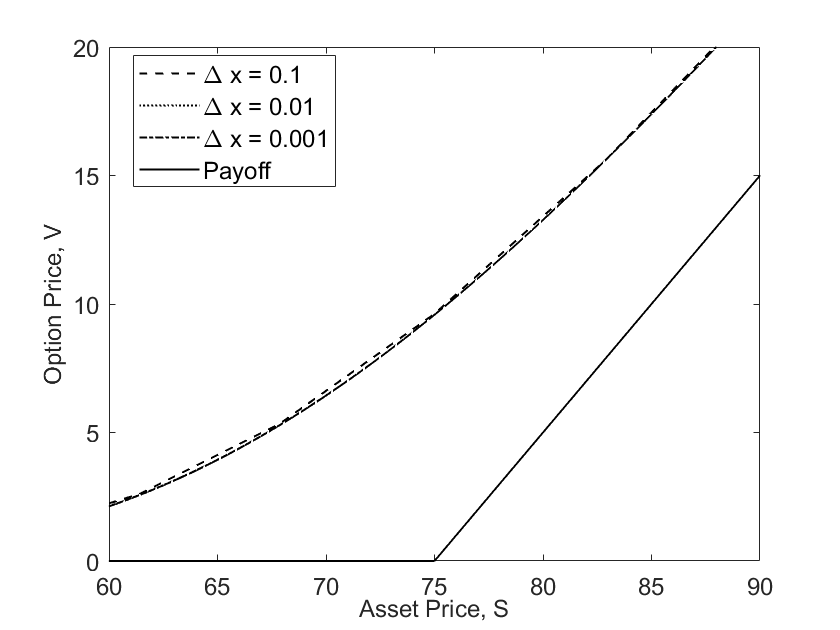}
\includegraphics[width=0.45\textwidth]{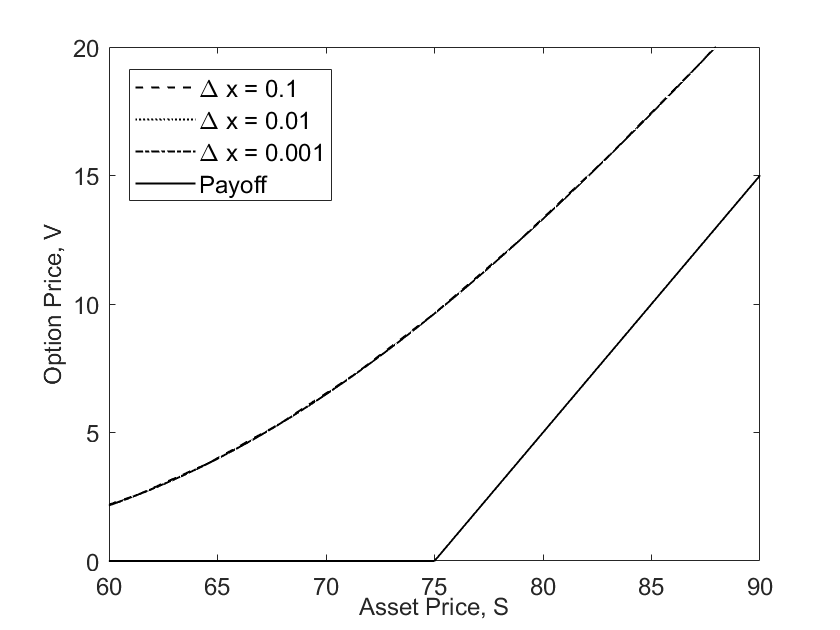}
\caption{Spatial mesh refinement effects of the FEM solutions, with $r = 0.1$, $\sigma = 0.2$, $K = 75$, $T = 1$, $C = 0.01$ and $M = 2$. Top: Group FEM with P1 (left) and P2 (right) element; Bottom: Numerical Quadrature with P1 (left) and P2 (right).} 
\label{fig:femrefinement}
\end{figure}

In Figure~\ref{fig:femrefinement}, we show numerical under uniform spatial-mesh refinement. As $\Delta x$ is reduced, the option price at $S = K$ tends to decrease. The decrease becomes insignificant as the mesh is refined from $\Delta x = 0.01$ to $0.001$. This may suggest convergence of the numerical solutions to a solution of the RAPM model, which need be analyzed theoretically.

\section{Conclusions} \label{sec:conclusion}

Several finite element-based models for approximations of the solutions of the RAPM Black-Scholes model were built in combination with some  Crank-Nicolson-type temporal scheme. Numerical examples demonstrated stable and accurate FEM solutions, which can be obtained by these models by  controlling the  spatial finite element sizes and the temporal step sizes. These numerical results compared favorably with those computed by finite difference schemes. The numerical results suggested that the numerical quadrature approach does not necessarily lead to a significantly better solution than the simple-to-construct group finite-element approach. Our finite element models  can be used as an effective alternative for numerical solutions of the RAPM model and can be adapted to solve similar models in option pricing.

\bibliography{main}

\newcommand{\noopsort}[1]{} \newcommand{\printfirst}[2]{#1}
  \newcommand{\singleletter}[1]{#1} \newcommand{\switchargs}[2]{#2#1}
\begin{thebibliography}{10}

\bibitem{Achdou05P}
Y.~Achdou and O.~Pironneau.
\newblock {\em Computational Methods for Option Pricing}.
\newblock SIAM, 2005.

\bibitem{Aichinger13B}
M.~Aichinger and A.~Binder.
\newblock {\em A Workout in Computational Finance}.
\newblock Wiley, 2013.

\bibitem{Almeida17CD}
R.M.P. Almeida, T.D.C. Chihaluca, and J.C.M. Duque.
\newblock Hermite finite element method for nonlinear {B}lack-{S}choles
  equation governing {E}uropean options.
\newblock In J.~Vigo-Aguiar et~al., editor, {\em Proceedings of the 17th
  International Conference on Computational and Mathematical Methods in Science
  and Engineering}, July 2017.

\bibitem{Anda11AS}
A.~Andalaft-Chacur, M.M Ali, and J.G. Salazar.
\newblock Real options pricing by the finite element method.
\newblock {\em Computers and Mathematics with Applications}, 61:2863--2873,
  2011.

\bibitem{AnkudinovaE08}
J.~Ankudinova and M.~Ehrhardt.
\newblock On the numerical solution of nonlinear {B}lack-{S}choles equations.
\newblock {\em Computers and Mathematics with Applications}, 56:799--812, 2008.

\bibitem{Barles98S}
G.~Barles and H.~Soner.
\newblock Option pricing with transaction costs and a nonlinear
  {B}lack-{S}choles equation.
\newblock {\em Finance and Stochastics}, 2(4):369--397, 1998.

\bibitem{scholesB73}
F.~Black and M.~Scholes.
\newblock The pricing of options and corporate liabilities.
\newblock {\em The Journal of Political Economy}, 81:637--654, 1973.

\bibitem{Company09JP}
R.~Company, L.~J\'odar, and J.-R. Pintos.
\newblock A numerical method for {E}uropean option pricing with transaction
  costs nonlinear equation.
\newblock {\em Mathematical and Computer Modelling}, 50(5--6):910--920, 2009.

\bibitem{Duering03FJ}
B.~D\"uring, M.~Fournier, and A.~J\"ungel.
\newblock High-order compact finite difference schemes for a nonlinear
  {B}lack-{S}choles equation.
\newblock {\em International Journal of Theoretical and Applied Finance},
  6(7):767--789, 2003.

\bibitem{Giles06C}
M.B. Giles and R.~Carter.
\newblock Convergence analysis of {C}rank–{N}icolson and {R}annacher
  time-marching.
\newblock {\em Journal of Computational Finance}, 9(4), 2006.

\bibitem{Golba13BA}
A.~Golbabai, L.V. Ballestra, and D.~Ahmadian.
\newblock Superconvergence of the finite element solutions of the
  {B}lack–{S}choles equation.
\newblock {\em Finance Research Letters}, 10:17--26, 2013.

\bibitem{Gulen19PS}
S.~Gulen, C.~Popescu, and M.~Sari.
\newblock A new approach for the black–scholes model with linear and
  nonlinear volatilities.
\newblock {\em Mathematics}, 7(8):760, 2019.

\bibitem{Hoggard94WW}
T.~Hoggard, A.~E. Whalley, and P.~Wilmott.
\newblock Hedging option portfolios in the presence of transaction costs.
\newblock {\em Advances in Futures and Options Research}, 7:21--35, 1994.

\bibitem{Janda05S}
M.~Janda\v{c}ka and D.~\v{S}ev\v{c}ovi\v{c}.
\newblock On the risk-adjusted pricing-methodology-based valuation of vanilla
  options and explanation of the volatility smile.
\newblock {\em Journal of Applied Mathematics}, 3:235--258, 2005.

\bibitem{liaoK09}
A.Q.M. Khaliq and W.~Liao.
\newblock High-order compact scheme for solving nonlinear {B}lack-{S}choles
  equation with transaction costs.
\newblock {\em International Journal of Computer Mathematics}, 86:1009--1023,
  2009.

\bibitem{Kratka98}
M.~Kratka.
\newblock No mystery behind the smile.
\newblock {\em Risk}, 9:67--71, 1998.

\bibitem{leland85}
H.~Leland.
\newblock Option pricing and replication with transactions costs.
\newblock {\em The Journal of Finance}, 40(5):1283--1301, 1985.

\bibitem{Linde09PS}
G.~Linde, J.~Persson, and L.~{von Sydow}.
\newblock A highly accurate adaptive finite difference solver for the
  {B}lack–{S}choles equation.
\newblock {\em International Journal of Computer Mathematics},
  86(12):2104--2121, 2009.

\bibitem{Marko08}
S.~Markolefas.
\newblock Standard {G}alerkin formulation with high order {L}agrange finite
  elements for option markets pricing.
\newblock {\em Applied Mathematics and Computation}, 195:707--720, 2008.

\bibitem{merton73}
R.~Merton.
\newblock Theory of rational option pricing.
\newblock {\em Bell Journal of Economics and Management Science},
  4(1):141--183, 1973.

\bibitem{Piron99H}
O.~Pironneau and F.~Hecht.
\newblock Mesh adaption for the {B}lack and {S}choles equations.
\newblock {\em East-West Journal of Numerical Mathematics}, 8(1), 1999.

\bibitem{Rann84}
R.~Rannacher.
\newblock Finite element solution of diffusion problems with irregular data.
\newblock {\em Numerische Mathematik}, 43:309--327, 1984.

\bibitem{vsevcovic2011analytical}
Daniel {\v{S}}evcovic, B~Stehl{\i}kov{\'a}, and K~Mikula.
\newblock Analytical and numerical methods for pricing financial derivatives.
\newblock {\em Nova Science, Hauppauge}, 2011.

\bibitem{Wei20EZ}
D.~Wei, Y.A. Erlangga, and G.~Zhumakhanova.
\newblock A finite element approach to the numerical solutions of {L}eland's
  model.
\newblock {\em submitted}, 2020.
\newblock available online at arxiv.org/abs/2010.13541.

\bibitem{Zhao16YW}
W.~Zhao, X.~Yang, and L.~Wu.
\newblock Alternating segment explicit-implicit and implicit-explicit parallel
  difference method for the nonlinear {L}eland equation.
\newblock {\em Advances in Difference Equations}, 103, 2016.
\newblock 18 pp, {DOI}: 10.1186/s13662-016-0823-5.

\end{thebibliography}
\bibliographystyle{plain}

\end{document}